\documentstyle[twocolumn,eqsecnum,aps]{revtex}
\def\be{\begin{equation}}
\def\ee{\end{equation}}
\def\bea{\begin{eqnarray}}
\def\eea{\end{eqnarray}}
\def\beas{\begin{eqnarray*}}
\def\eeas{\end{eqnarray*}}

\begin{document}
\draft
\title{A Study of Heavy-Light Mesons on the Transverse Lattice}
\author{Matthias Burkardt and Sudip K. Seal}
\address{Department of Physics\\
New Mexico State University\\
Las Cruces, NM 88003-0001\\U.S.A.}
\maketitle
\begin{abstract}
We present results from a study of meson spectra and structure
in the limit where one quark is infinitely heavy. The calculations,
based on the framework of light-front QCD formulated on
a transverse lattice, are the first non-perturbative studies of B-mesons in light-front
QCD. We calculate the Isgur-Wise form factor, light-cone 
distribution amplitude, the heavy-quark parton distribution function and
the leptonic decay constant of B-mesons.
\end{abstract}
\narrowtext
\section{Introduction}
Heavy flavor physics has become one of our most promising tools in
searching for physics beyond the standard model. 
However, despite the use of heavy quark symmetry and heavy quark effective 
field theory, many uncertainties in extracting standard model parameters
from data on heavy meson decays arise from the often unknown 
nonperturbative hadronic matrix elements for these decays.
For example, in order to extract the CKM matrix matrix element 
$\left|V_{bc}\right|$ from semileptonic $\bar{B}\rightarrow D^*$ transitions,
it is necessary to extrapolate to the point $v\cdot v^\prime=1$.
Even in the heavy quark limit, this involves the Isgur-Wise form factor,
which is not determined by symmetry. For this and similar applications,
it is essential to have reliable predictions for hadronic
matrix elements that multiply interesting standard model parameters in the 
physical amplitudes.\footnote{For a recent review, see Ref. \cite{neubert}.}

Euclidean lattice gauge theories have played an important role for calculating 
some of the relevant nonperturbative matrix elements. However, especially for
matrix elements that involve a large velocity transfer as well as matrix
elements that probe light-like correlation functions (such as the $B$-meson
distribution amplitude or the parton distribution functions, which are 
relevant for inclusive semileptonic decays), it would be much more natural to 
describe the nonperturbative hadronic physics in the light-front (LF)
framework (or infinite momentum frame), i.e. in a Hamiltonian framework
where $x^+\equiv x^0+x^3$ is `time'.

What distinguishes the LF framework from all other
formulation of QCD is that the observables, which are dominated by
light-like correlation functions, become equal `time' correlation functions
and hence have a very direct and
physical connection to the microscopic degrees of freedom in terms of which the
Hamiltonian is constructed \cite{mb:adv}. 
Because of this unique feature, it should be
much easier in this framework to gain a physical understanding between
experiment and phenomenology on the one hand and the
underlying QCD dynamics on the other.

Of course, just like in any other approach to QCD, it is necessary to
regularize
both UV and IR divergences before one can even attempt to perform
nonperturbative calculations.
The transverse lattice \cite{bardeen} is an attempt to combine advantages
of the LF and lattice formulations of QCD.
In this approach to LF-QCD the time and one space direction (say $x^3$)
are kept continuous, while the two `transverse' directions
${\bf x}_\perp \equiv (x^1,x^2)$ are discretized.
Keeping the time and $x^3$ directions continuous has the advantage of
preserving manifest boost invariance for boosts in the $x^3$ direction.
Furthermore, since $x^\pm = x^0 \pm x^3$ also remain continuous,
this formulation still allows a canonical LF Hamiltonian approach.
On the other hand, working on a position space lattice in the transverse
direction allows one to introduce a gauge invariant cutoff on
$\perp$ momenta --- in a manner that is similar to Euclidean or
Hamiltonian lattice gauge theory.

In summary, the LF formulation has the advantage of utilizing degrees of
freedom that are very physical since many high-energy scattering observables
(such as deep-inelastic scattering cross sections) have very simple
and intuitive interpretations as equal LF-time ($x^+$) correlation
functions.
Using a gauge invariant (position space-) lattice cutoff in the $\perp$
direction within the LF framework has the advantage of being able to avoid
the notorious $1/k^+$ divergences from the gauge field in LF-gauge which
plague many other Hamiltonian  LF approaches to QCD \cite{mb:korea}.

The hybrid (continuous along with discrete) treatment  of the longitudinal/transverse
directions implies an analogous hybrid treatment of the longitudinal versus transverse
gauge field: the longitudinal gauge field degrees of freedom
are the non-compact $A^\mu$ while the transverse gauge degrees of freedom are
compact link-fields. Each of these degrees of freedom depend on two
continuous
($x^\pm$) and two discrete (${\bf n}_\perp$) space-time variables, i.e. from
a formal point of view the canonical
transverse lattice formulation is equivalent to a
large number of coupled $1+1$ dimensional gauge theories
(the longitudinal gauge fields at each ${\bf n}_\perp$) coupled to nonlinear
$\sigma$ model degrees of freedom (the link fields) \cite{paul}.

This formalism has been previously applied to pure-glue QCD 
(numerical results for the static
quark-antiquark potential can for example be found in Ref. \cite{mb:bob} 
and for glueball spectra in Ref. \cite{brett}).
More recent applications of the color dielectric transverse lattice formulation
to mesons at large $N_C$ can be found in Refs. \cite{sd:meson,sudip,studs}, where more details regarding the transverse lattice approach approach can be found.

In this paper, we make the first attempt to apply the $\perp$ lattice, used in the above light 
meson calculations, to mesons where one of the quarks is infinitely heavy.
The paper is organized as follows.  We briefly describe the Hamiltonian, 
the approximations used to solve it in section 2. 
Numerical results for a number of important hadronic matrix elements  are presented in section 3.

\section{The Hamiltonian}
The transverse lattice Hamiltonian for Wilson fermions
that forms the basis of our work was
first constructed in Ref. \cite{mb:hala} and was previously applied to
mesons containing only light quarks in Refs. \cite{sd:meson,studs}, and we 
refer the reader to these works for more details. 
Here we restrict ourselves to a more qualitative
description of the degrees of freedom that enter the Hamiltonian and some
of the details of how heavy quarks are incorporated in this framework.

The degrees of freedom
that enter the Hamiltonian are quark and antiquark creation operators at
each site and link-fields (which, in the spirit of the color-dielectric
approach \cite{brett}, we take as general  matrix fields) on the transverse
links connecting the sites. For simplicity, we work in the large $N_C$ 
limit. The first approximation that we use is the same
light front Tamm-Dancoff truncation of the Fock state expansion that was used
in Refs. \cite{sd:meson,studs}, namely, a truncation upto and
including the 3-particle sector. After reminding oneself that the fermion
degrees of freedom (quarks and antiquarks)
always occupy the lattice (transverse) sites and the gauge degrees of freedom
(link fields) reside
on the lattice spacings, it is easy to understand the following two direct
consequences of the
aforementioned truncation:
\begin{itemize}
\item The 2-particle states consist of a quark and an antiquark sitting on
the same lattice site.
\item The 3-particle states consist of a quark and an antiquark separated by
{\bf at most} one link field.
\end{itemize}

The Hamiltonian for the light quarks and link fields in this model
contains the following 6 parameters
\begin{itemize}
\item kinetic masses for the light quarks in the 2 and 3 particle Fock sector
\item helicity flip and non-flip hopping terms for the light quarks
\item longitudinal gauge coupling
\item kinetic mass for the link field
\end{itemize}
which were determined in Ref. \cite{studs} by imposing rotational invariance,
fixing the physical string tension in $\perp$ lattice units in both the
longitudinal and transverse direction and the $\pi$ and $\rho$ masses.

For the heavy quarks one has a variety of options. Conceptually, it would be cleanest to treat them as infinitely heavy fixed sources, using the fixed
source formalism outlined in Ref. \cite{mb:bob}. In such an approach, no new
parameters would enter the Hamiltonian. In this work, we adopted a slightly different procedure to implement the heavy quarks: we treated them as finite mass quarks, but did not allow the heavy quarks to propagate in the transverse
direction --- only propagation in the longitudinal direction is allowed.
The advantage if this hybrid procedure is that we could utilize 
computer code that was used to describe light-light mesons and the heavy quark
limit results were then obtained by extrapolation.
\footnote{Note that although this procedure produces the correct infinite quark mass
results through extrapolation, we cannot use the results to study finite $M_b$
corrections, since the dynamics of finite $M_b$ is not properly described
(no transverse hopping!).}
Treating heavy quarks consistently as finite quark mass heavy quarks would have the advantage that one could study $1/M_b$ corrections to the heavy quark limit.
However, since this would not only require developing an analog of the NRQCD
procedure on the LF, but also a determination of a whole new set of parameters
(the coefficients of the hopping terms for the heavy quark in the Hamiltonian),
we preferred to focus only on the $M_b\rightarrow \infty$ limit, where no new
parameters are introduced.

\section{Numerical Results}
The observables that we studied are the Isgur Wise form factor, the 
decay constant $f_B$, the $B$-meson light cone distribution amplitude and
the (light-quark) parton distribution function in a $B$-meson.
For all these observables, we performed calculations at finite
kinetic mass $M_b$ (but no $\perp$ hopping) and then extrapolated to
$M_b\rightarrow \infty$. Any error bars/bands that we show for these
observables reflect uncertainties arising from this extrapolation.

In each of these calculations, we diagonalized the $\perp$ lattice 
Hamiltonian in a truncated basis, using continuous basis functions.
As output of these calculations we obtained the wave functions
(Fock space amplitudes) in the two and three particle Fock components
$\psi_s(x)$ and $\psi_s(x,y)$ where the arguments in the two particle Fock 
component refer to the momentum fraction carried by the light quark and 
in the three particle Fock component refers to the momentum fractions
carried by the light and heavy quark. The discrete index $s$ labels the
spin configurations ($4$ in the two particle Fock component) and the
spin/orientation components 
($4\times 4=16$ in the three particle Fock component).

In the following, we will use numerical results for these wave functions
to calculate some important hadronic matrix elements for these
heavy-light systems.

One of the most important hadronic observables in $B$-physics is the
Isgur-Wise form factor. Heavy quark symmetry predicts that 
the matrix element of a heavy quark current operator in heavy-light mesons
is described by a universal form factor
\be 
\langle B(p^\prime) | \bar{b}\gamma^\mu b |B(p)\rangle
= M_B \left( v^\mu + v^{\prime \mu}\right) F(v\cdot v^\prime)
\ee
Although this form factor
does still depend on the quantum numbers of the light 
degrees of freedom, it becomes independent
of the heavy quark mass as $M_Q\rightarrow \infty$.
Knowledge of this form factor is very useful for determining the CKM
matrix element $V_{bc}$ from semi-leptonic decays $\bar{B} \rightarrow
D^* e\nu_e$ because it allows one to extrapolate decay data
to the zero recoil point $v\cdot v^\prime\rightarrow 1$. 
Within the context of the transverse lattice, the form factor is most 
conveniently extracted from the $+$ component of the current operator,
for which we find
\be
F(v\cdot v^\prime) = F^{(2)} (v\cdot v^\prime) + F^{(3)} (v\cdot v^\prime)
,
\label{eq:ff}
\ee
where 
\bea
F^{(2)} (x)&=& \frac{2}{2-x}\sum_s \int^1_x dz
\psi_s(z)\psi_s^*\left(\frac{z-x}{1-x}\right)\\
F^{(3)} (x)&=&\frac{2}{2-x}\frac{1}{\sqrt{1-x}}\sum_s
\int_x^1dz\\&&\int_0^{1-z}dw\psi_s(z,w)
\psi_s^*\left(\frac{z-x}{1-x},\frac{w}{1-x}\right)
\eea
are the contributions from the two and three particle Fock components 
respectively and $\sum_s$ refers to the summation over the discrete
spin and orientation label for the meson.
The velocity transfer is related to light-cone plus momentum transfer $x$
through the relation 
$v\cdot v^\prime= \frac{1}{2}\left(1-x+\frac{1}{1-x}\right)$.
Note that the normalization condition $F(v\cdot v^\prime=1)=1$ 
is automatic
in this framework because of the normalization condition for the wave 
function. Note also that terms that are off-diagonal in Fock space, and which
in general have to be included when calculating current matrix elements
when $q^+\equiv {p^+}^\prime -p^+$ is nonzero \cite{stan}, are suppressed
in the limit $M_b\rightarrow \infty$ (at least when considering the 
`good' component).

In the numerical calculations we evaluated Eq. (\ref{eq:ff}) for a variety
of heavy quark masses and extrapolated to $M_b\rightarrow \infty$.
Results for the extrapolated form factor are depicted
in Fig. \ref{fig:form}, where we multiplied the form factor with
$V_{bc}=0.0424$ \cite{exp} 
in order to be able to compare to experimental data
\cite{exp}. Overall, our calculated form factor is in fairly good agreement
with the data, although the experimental error bars are quite large.
We should perhaps point out that if we include the data at larger
values of $v\cdot v^\prime$ into account then we would obtain a slightly
better fit to the data for values of $V_{bc}$ in the range
$V_{bc}=0.0415- 0.0420$. The functional form of the fit to our calculated 
form factor as presented in Fig. \ref{fig:form} is:

\begin{figure}
\unitlength1.cm
\begin{picture}(15,7.0)(-8.0,0.0)
\includegraphics{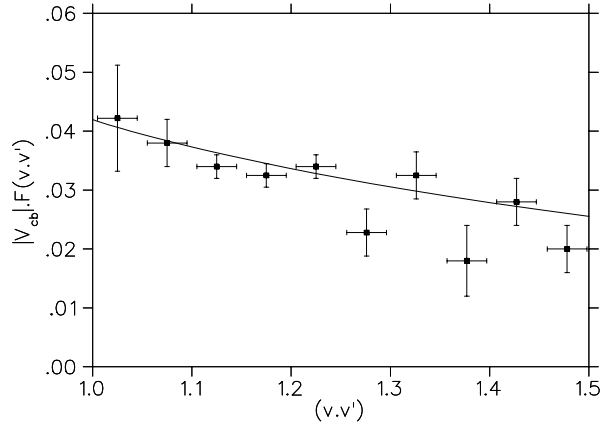}
\end{picture}
\caption{Isgur-Wise form factor (\ref{eq:ff}) in the heavy quark limit.
Uncertainties from the $M_b\rightarrow \infty$ limit were very small for this 
observable. The experimental data is from Ref. [14].}
\label{fig:form}
\end{figure}

\bea
F(v.v')&=&1-1.398(v.v'-1) +1.722(v.v'-1)^2\nonumber\\
&&
-0.203(v.v'-1)^3 
-1.824(v.v'-1)^4
\eea

Although no experimental data exists on 
\be
\langle 0 |\bar{b}\gamma^\mu \gamma_5u|B\rangle = p^\mu f_B,
\ee
it would be very useful to have predictions for this observable because
of its role in B-meson phenomenology.

Up to an overall scale factor 
which involves the $\perp$ lattice spacing,
$f_B$ can be conveniently expressed in terms of the two particle 
wave function on the transverse lattice \cite{gross}
\be
f_B=\frac{2}{a_\perp}\sqrt{\frac{N_C}{\pi}} \int_0^1 dx \psi_{+-}(x) .
\ee
We find (in the heavy quark limit) for a value of $M_b=5 GeV$ and $a_\perp\approx 0.5fm$ \cite{studs}
\be
f_B = 238\pm 20\, MeV \label{eq:ratio}
\ee
where the error bar reflects only the uncertainty in our numerical
extrapolation to the heavy quark limit. A much larger systematic
uncertainty in Eq. (\ref{eq:ratio}) arises due to our Fock space
truncation. We expect that including more higher Fock components
will significantly reduce the 2-particle Fock space amplitude and
hence also the decay constant, but it is very difficult to estimate
the size of the reduction.

\begin{figure}
\unitlength1.cm
\begin{picture}(15,7.0)(-8.0,0.0)
\includegraphics{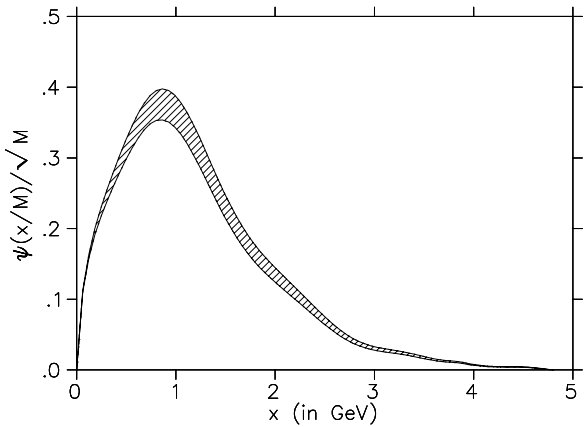}
\end{picture}
\caption{B-meson distribution amplitude in the heavy quark limit.
Systematic uncertainties arising from the numerical extrapolation are
indicated by the shaded area.}  
\label{fig:wf}
\end{figure}

In addition to the decay constant, we also studied the light-cone distribution
amplitude itself. In the LF approach, the twist-2
distribution amplitude
is given by the wave function in the two particle Fock space sector with the
quark and the antiquark at the same transverse position. On the transverse
lattice, the distribution amplitude, which plays a role in large momentum
transfer processes, is thus conveniently expressed in terms of
the two particle Fock component wavefunction $\psi(x)$
\be
\phi(x)= N \psi(x),
\ee
where $N$ is a normalization constant.
The scaling behavior of $\phi(x)$ is such that
\be
\phi_\infty(x)\equiv \lim_{M_b\rightarrow \infty}
\frac{1}{\sqrt{M_b}}\phi(x/M_b)
\ee
scales in the heavy quark limit. Numerical results for $\phi_\infty(x)$
are shown in Fig. \ref{fig:wf}. Although a precise determination of the
power law behavior near $x\rightarrow 0$ was 

not possible, we found
that our numerical results are consistent with a power close to $0.5$
below $x$=1 GeV. For the lowest moments, we find

\be
M_{-1}\equiv \frac{\int_0^\infty \frac{dz}{z}\phi_\infty(z)}
{\int_0^\infty dz\phi_\infty(z)}=1.509\,GeV^{-1}
\label{eq:moment1}
\ee
\be 
M_1\equiv
\frac{\int_0^\infty dzz\phi_\infty (z)}{\int_0^\infty dz\phi_\infty(z)}=
1.223\, GeV.
\label{eq:moment2}
\ee
From Eqs. (\ref{eq:moment1}) and (\ref{eq:moment2}), it is clear
that $\bar{\Lambda}=\frac{3}{4}M_1\approx 0.9\, GeV $ \cite{lambda}
in our calculations turns out to be very large.
This result is consistent with the `binding energies' $M_B-M_b$
that we obtain
and reflect the fact that the bare (unscreened) longitudinal string
tension is very large. We expect that including higher Fock states
will lead to a significant lowering of the longitudinal string tension
and hence also $\bar{\Lambda}$.

\begin{figure}
\unitlength1.cm
\begin{picture}(15,7.0)(-8.0,0.0)
\includegraphics{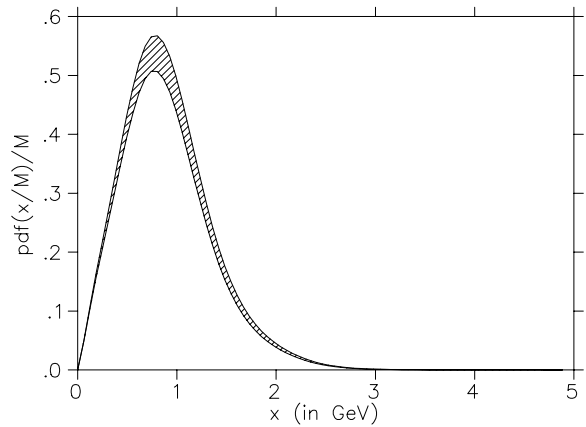}
\end{picture}
\caption{Light-cone momentum distribution (\ref{eq:f}).
The shaded area reflects
uncertainties arising from the numerical extrapolation.}
\label{fig:f}
\end{figure}

The light-cone momentum distribution of the heavy quark inside a B-meson
is relevant for
the invariant mass distribution in inclusive decays of heavy-light
systems because it determines the amount of energy that is available for
the weak decay products of the heavy quark.
Of course, to leading order in $1/M_b$ the {\it b} quark carries 100\% of the
B-meson's momentum in the infinite momentum frame.
We thus consider the first correction to this trivial result, i.e.

\be
f_b(z) = \lim_{M_b\rightarrow \infty} b\left( 1-\frac{z}{M_b}\right)
\label{eq:f}
\ee

where $b(x)$ is the light-cone momentum distribution of the
$b$-quark. Results are shown in Fig. \ref{fig:f}. Since the 2 particle Fock
component is the dominant component for $B$-mesons in our calculations,
$f_b(z)$ is to a good approximation described by $\left|\psi_\infty(z)\right|^2$
(Fig. \ref{fig:wf}).

\section{Summary and Outlook}
We have performed the first non-perturbative study of mesons containing one
heavy and one light quark in light front QCD based on the transverse lattice
fomulation.
The parameters that appear in the Hamiltonian 
for a heavy-light system are the same that appear also
in the Hamiltonian for light-light mesons. Since we determined the latter
in a previous work \cite{studs} using constraints from rotational invariance,
we were thus able to make parameter free predictions for a variety of
heavy meson observables.

Specifically, we studied the Isgur-Wise form factor, decay constant,
light-cone distribution amplitude and light-cone momentum distribution function
for $B$ mesons in the limit $M_b\rightarrow \infty$.
Our results for the form factor are consistent with experimental data and
favor a CKM matrix element $V_{bc}$ in the range $0.0415-0.0420$.

Our main sources of systematic errors are the Fock space truncation
(in combination with a truncation of the transverse lattice Hamiltonian
\cite{sd:meson}), the truncation of the Hilbert space within each Fock
component and the extrapolation to the $M_b\rightarrow \infty$ limit.
The uncertainties from Hilbert space truncations are well under control
since we used basis function techniques and extrapolated the results
in the dimension of the Hilbert space. The $M_b\rightarrow \infty$ are also
well under control since we studied a number of different $M_b$ values.
The main source of systematic errors in our calculations
arise from the Fock space truncation and it is difficult to estimate the
size of the corrections from higher Fock components without actually
including them. Although we expect little corrections from higher Fock components for the form factor at small velocity transfers (the
main effect of higher Fock component would be to provide the quarks with an additional intrinsic form factor which would only be probed at large 
momentum transfers), higher Fock components probably have a large effect
on the decay constant which is proportional to the amplitude to find the
meson in the two particle Fock component.
We did not include higher Fock components in this work because,
for consistency, this requires inclusion of a number of new operators in the $\perp$ lattice Hamiltonian --- a project that would have been beyond the intended scope of this one.
Nevertheless, it is clear that one possible future extension of this work 
should be to take higher Fock components into account. 

There are many other natural extensions of this work. For example, it would
be very useful to repeat this study for $B_s$ mesons. While the $u\bar{b}$ 
meson
calculations that we performed here could make use of light-quark parameters
determined in previous works \cite{studs,sd:meson}, such a study involving
$s$ quarks would first require determination of hopping parameters for
$s$ quarks using techniques similar to the used in Refs. \cite{studs,sd:meson}.

Although we performed the numerical calculations for a variety of heavy
quark masses, we did not allow the heavy quarks to propagate in the transverse 
directions because this would have forced us to introduce (and determine)
transverse hopping parameters. The advantage of this approach was that we were
able to study the $M_b \rightarrow \infty$ limit without introducing any new
parameters. The disadvantage is that we cannot use the results to study
the $1/M_b$ to the heavy quark observables, which would be extremely
useful for the analysis of experimental data, especially those involving 
charm quarks. 

\noindent {\bf Acknowledgments}
This work was supported by a grant from DOE (FG03-95ER40965) and through
Jefferson Lab by contract DE-AC05-84ER40150 under which the Southeastern
Universities Research Association (SURA) operates the Thomas Jefferson
National Accelerator Facility.

\appendix


\begin{references} \bibitem{neubert} M. Neubert, lecture notes,
hep-ph/0012204.  \bibitem{Brodsky:1998de} S.~J.~Brodsky, H.~Pauli and
S.~S.~Pinsky, 
Phys.\ Rept.\ {\bf 301}, 299 (1998)
[hep-ph/9705477].     
\bibitem{mb:adv} M. Burkardt, Advances Nucl. Phys. {\bf 23},{1}
{(1996)}.
\bibitem{bardeen} W.A. Bardeen, R.B. Pearson, and E. Rabinovici,
Phys. Rev. D{\bf 21}, 1037 (1980)
\bibitem{mb:korea} M. Burkardt, invited talk given at ``11th International
Light-Cone School and Workshop'', Eds. C. Ji and D.-P. Min,
hep-th/9908195.
\bibitem{paul} Griffin, P.A., in {\it Theory of Hadrons and Light Front QCD}, 
ed. S.D. Glazek, World Scientific, 1995, p. 240, hep-th/9410243. 
\bibitem{mb:bob} M. Burkardt and B. Klindworth, Phys. Rev. D{\bf 55}, 1001
(1997)
\bibitem{brett} S. Dalley and B. vande Sande,
Phys. Rev. D{\bf 59}, (1999); Phys. Rev. Lett.{\bf 82}, 1088 (1999);
Nucl. Phys. B (Proc. Suppl.){\bf 83}, 116 (2000);
Phys. Rev. D{\bf 62},(2000)
\bibitem{sd:meson} S. Dalley,
Nucl. Phys. B (Proc. Suppl.){\bf 90}, 227 (2000); 
S. Dalley, hep-ph/0101318.
\bibitem{sudip} S.K. Seal and M. Burkardt, Nucl. Phys. B (Proc. Suppl.) 90,
233(2000)
\bibitem{studs} M. Burkardt and S.K. Seal, hep-ph/0102245.
\bibitem{mb:hala} M. Burkardt and H. El-Khozondar,
Phys. Rev. D{\bf 60}, (1999)
\bibitem{stan} S.J. Brodsky and D.S. Hwang, Nucl.\  Phys.\  B\
{\bf 543}, 239 (1999).
\bibitem{exp} CLEO Collaboration, hep-ex/0007052.
\bibitem{gross} C.G.Callen, N.Coote, and D.J.Gross, 
Phys.\ Rev.\ D\ {\bf 13}, 1649 (1976).
\bibitem{lambda} M. Beneke and T. Feldmann, Nucl.\ Phys.\ B\ {\bf 592}, 3
(2001).
\end{references}
\end{document}